\documentclass[11pt, fceqn]{article}
 \usepackage{amsfonts}
 \usepackage{graphicx}
 \usepackage{flafter}
 \usepackage{amssymb,graphics,graphicx,subfigure,caption2,rotating}
 \usepackage{amsmath, latexsym}
\usepackage[amsmath,thmmarks]{ntheorem}
 \usepackage[top=3.5cm]{geometry}
\begin{document}
\date{}
\title{Optimization of ultrafine entanglement witnesses}
\author{Shu-Qian Shen$^{1,2}$\thanks{Email: sqshen@upc.edu.cn}, Ti-Run Xu$^1$, Shao-Ming Fei$^{2,3}$\thanks{Email: feishm@cnu.edu.cn}, Xianqing Li-Jost$^{2}$\thanks{Email: xli-jost@mis.mpg.de},  Ming Li$^1$\thanks{Email: liming3737@163.com}\\
{\small{\it $^1$College of Science, China University of Petroleum, 266580 Qingdao, P.R. China}}\\
{\small{\it $^2$Max-Planck-Institute for Mathematics in the Sciences, 04103 Leipzig, Germany}}\\
{\small{\it $^3$School of Mathematical Sciences, Capital Normal University, 100048 Beijing, P.R. China}}}

\maketitle
\begin{abstract}
The ultrafine entanglement witness, introduced in [F. Shahandeh, M. Ringbauer, J.C. Loredo, and T.C. Ralph, Phys. Rev. Lett.  \textbf{118},  110502 (2017)], can seamlessly and easily improve any standard entanglement witness. In this paper, by combining the constraint and the test operators, we rotate the hyperplane determined by the test operator and improve further the original ultrafine entanglement witness. In particular, we present a series of new ultrafine entanglement witnesses, which not only can detect entangled states that the original ultrafine entanglement witnesses cannot detect, but also have the merits that the original ultrafine entanglement witnesses have.
\end{abstract}

\section{Introduction}
Quantum entanglement \cite{Einstein1935,Schrodinger1935} is one of the distinguished features of quantum mechanics. It provides a necessary resource in most quantum information processing \cite{Nielsen2010,survey}. A natural problem is how to detect and quantify entanglement. Unfortunately, it is still an open problem of determining whether any given quantum state is entangled or separable. This difficult problem has motivated the development of a variety of separability criteria in the last decades \cite{survey}. However, a large amount of separability criteria need the quantum state tomography \cite{tomography} to reconstruct the quantum states, which is impractical for high-dimensional and multipartite quantum states in experimental implementations \cite{qudittomogrphy}.

Without the need of quantum state tomography, entanglement witness \cite{Horodecki1996,Terhal2000} is an economical tool for entanglement detection. A Hermitian operator is called an entanglement witness (EW) if and only if it has non-negative expectation values over all separable states, and has negative expectation values at least for one entangled state \cite{Terhal2000}. It was shown in \cite{Horodecki1996,Terhal2000} that any entangled state can be detected by some entanglement witnesses. From the viewpoint of convex sets, any entanglement witness can be seen as a hyperplane that distinguishes some specific entangled states from the rest ones. In the last years, various  constructions of EWs have been studied; see \cite{Chruscinski2014} for a comprehensive survey. In particular, Sperling and Vogel \cite{Sperling2009} proposed a general form of entanglement witness, which is called the standard EW (SEW) \cite{Shahandeh2017}.

Recently, Shahandeh et. al. \cite{Shahandeh2017} gave an innovative method, i.e., the ultrafine EW (UEW), for entanglement detection, which can easily improve any SEW. In comparison with the SEW, the UEW has mainly four merits. Firstly, the eigenvectors corresponding to the maximum eigenvalue of test operators need not be entangled. Secondly, the UEW removes the need of the witness decomposition into local  observables. Thirdly, some UEWs can be easily implemented by  using a simple measurement device for each party. Finally, the UEW can be used to improve any SEW.

In this paper, by rotating the hyperplane determined by the linear combination of the constraint and the test operators, we present some new ultrafine entanglement witnesses. The optimal ultrafine entanglement witness among them is studied. In addition, a detailed example is used to illustrate the optimization process.

The paper is organized as follows. After introducing UEWs in Section 2, we give the new UEWs by rotating the corresponding hyperplane in Section 3. And then the optimal UEW is also investigated. Some concluding remarks are given
in Section 4.

\section{SEWs and UEWs}
We first introduce some preliminaries about SEWs and UEWs in \cite{Sperling2009,Shahandeh2017}. Let $H_A$ and $H_B$ be arbitrary finite dimensional Hilbert spaces. We denote by $\mathcal{S}_{sep}$ the set of separable states in $H_A\otimes H_B$. If an EW $W$ acting on $H_A\otimes H_B$ can be written into the following general form \cite{Sperling2009,Lewenstein2000}:
\begin{equation}
\label{generW} W(L)=g_s(L) I-L,
\end{equation}
where $I$ is the identity operator, $L$ is called the test operator, and $g_s(L)$ is the supremum of expectation values taken over all separable pure states $|a,b\rangle$:
$$
g_s(L)=\text{sup}\{\langle a,b|L|a,b\rangle\},
$$
then this EW is said to be a finest EW (FEW). Thus, there exists a state $|a_0,b_0\rangle$ satisfying $\langle a_0,b_0|W(L)|a_0,b_0\rangle=0$, which is called the optimal state in $\mathcal{S}_{sep}$ to the test operator $L$. Clearly, $W(L)$ can detect entanglement if and only if its eigenvectors corresponding to the maximum eigenvalue of $L$ are entangled.

Another Hermitian operator $C\neq L$  is a physical observable. The whole  state space is cut into two half-spaces by $C$:
\begin{equation*}
\mathcal{S}_c=\{\rho|\text{Tr}(\rho C)\le c\},~~~ \mathcal{S}_{\tilde{c}}=\{\rho|\text{Tr}(\rho C)\ge c\},
\end{equation*}
where $c$ is any real-valued constant. Here, $C$ is called the constraint operator. Further, we define the following two sets:
\begin{equation}
\label{twosets}\mathcal{S}_{sep:c}=\mathcal{S}_{sep}\cap \mathcal{S}_{c},~~~ \mathcal{S}_{sep:\tilde{c}}=\mathcal{S}_{sep}\cap \mathcal{S}_{\tilde{c}}.
\end{equation}
If one of the sets $ \mathcal{S}_{sep:c}$ and $ \mathcal{S}_{sep:\tilde{c}}$ is empty, then the other set coincides with the set of separable states $\mathcal{S}_{sep}$. In this case, the UEW constructed in \cite{Shahandeh2017} reduces to the standard EW. \emph{Hence, we always assume that the two sets $ \mathcal{S}_{sep:c}$ and $ \mathcal{S}_{sep:\tilde{c}}$ are both non-empty.}

Based on the test operator $L$, the UEWs $W_c(L)$ and $W_{\tilde c}(L)$  \cite{Shahandeh2017} are then defined as
\begin{equation}\label{p1}
W_c(L)=p_c(L)I-L,~~~ W_{\tilde c}(L)=p_{\tilde c}(L)I-L,
\end{equation}
where
\begin{equation}\label{p2}
p_c(L)=\text{sup}\{\text{Tr}(L\rho)|\rho\in \mathcal{S}_{sep:c}\},~~~ p_{\tilde c}(L)=\text{sup}\{\text{Tr}(L\rho)|\rho\in \mathcal{S}_{sep:\tilde{c}}\}.
\end{equation}
Thus,  the ultrafine entanglement witness \cite{Shahandeh2017} states that a given state $\rho$ is entangled if
\begin{align}
\label{uew}\text{Tr}(C\rho)\le c \wedge \text{Tr}(W_c(L)\rho)<0~\text{or}~\text{Tr}(C\rho)\ge c \wedge \text{Tr}(W_{\tilde{c}}(L)\rho)<0.
\end{align}

Due to \cite[Lemma 1]{Shahandeh2017}, the optimal state from the FEW $W(L)$ must be an optimal state from either $W_c(L)$ in $\mathcal{S}_{sep:c}$ or $W_{\tilde c}(L)$ in $\mathcal{S}_{sep:\tilde{c}}$. Thus, one of $W_c(L)$ and $W_{\tilde c}(L)$ must coincide with $W(L)$, while the other can detect some entangled states that $W(L)$ cannot detect. Hence, the entanglement condition (\ref{uew}) is always better than that of FEW $W(L)$. \emph{Thus, without loss of generality, in what follows we always assume that  the optimal state  in $\mathcal{S}_{sep}$ to the test operator $L$ belongs to $\mathcal{S}_{sep:\tilde c}$.}

%
\section{Optimization of UEWs}
In order to carry out rotation,  we define a linearly combined operator from the constraint operator $C$ and the test operator $L$:
\begin{align*}
N_{\alpha}&=\alpha C+(1-\alpha)L,~~~ \forall 1>\alpha\in \mathbb{R},
\end{align*}
which can be seen as a new test operator instead of $L$. Based on $N_{\alpha}$ and
the definition of UEW, we can get a series of ultrafine entanglement witnesses. For convenience, we call them $\alpha$-UEWs.
\\
\\
\textbf{$\alpha$-UEWs}: \emph{For a given constraint value $c$ and any real parameter $\alpha<1$, a state $\rho$ is entangled if
\begin{align}
\label{uew1}\emph{Tr}(C\rho)\le c \wedge \emph{Tr}(W_c(N_{\alpha})\rho)<0~\text{ or }~ \emph{Tr}(C\rho)\ge c \wedge \emph{Tr}(W_{\tilde{c}}(N_{\alpha})\rho)<0,
\end{align}
where $W_c(N_{\alpha})$ and  $W_{\tilde{c}}(N_{\alpha})$ are similarly defined as in \emph{(\ref{p1})} and \emph{(\ref{p2})}, with $L$ replaced by $N_{\alpha}$.}
\\

Clearly, for the case $\alpha=0$, the $\alpha$-UEW $W_c(N_{\alpha})$ reduces to the UEW $W_c(L)$.
We shall study the efficiency of the UEW $W_c(N_{\alpha})$ when the parameter $\alpha$ gets smaller from $1$ to $-\infty$. The following lemmas are needed.
\\
\\
\textbf{Lemma 3.1.} \emph{For a given constraint value $c$ and a real constant $\alpha<1$, if the optimal pure states in $\mathcal{S}_{sep}$ to the test operators $L$ and $N_{\alpha}$ both belong to $\mathcal{S}_{sep:\tilde c}$, then the optimal pure state $\sigma$  in $\mathcal{S}_{sep:c}$  to the test operator $L$  is also an optimal state in $\mathcal{S}_{sep:c}$ to the test operator $N_{\alpha}$ with $\emph{\text{Tr}}(\sigma N_{\alpha})=\alpha c+(1-\alpha)p_c(L)$}.
\\
\\
\textbf{Proof.}  From \cite[Theorem 1]{Shahandeh2017}, there exists a pure state $\sigma\in \mathcal{S}_{sep:c}$ satisfying
$\text{Tr}(\sigma C)=c, \text{Tr}(\sigma L)=p_c(L)$, and then
\begin{equation}
\label{Nalpha}
\text{Tr}(\sigma N_{\alpha})=\alpha c+(1-\alpha)p_c(L).
\end{equation}
 Define
$h_{\alpha:c}=\sup\limits_{\rho\in \mathcal{S}_{sep:c}}\{\text{Tr}(\rho N_{\alpha})\}$.
In order to show that the state $\sigma$ is the optimal state in $\mathcal{S}_{sep:c}$ to the test operator $N_{\alpha}$, we only need to prove
\begin{align}
\label{conclusion} h_{\alpha:c}=\alpha c+(1-\alpha)p_c(L).
\end{align}
In fact, from (\ref{Nalpha}) it is easy to deduce
\begin{align}
\label{halpha} h_{\alpha:c}\ge \alpha c+(1-\alpha)p_c(L).
\end{align}
On the other hand, suppose that $|a,b\rangle $ in $\mathcal{S}_{sep:c}$  is optimal to
the test operator $N_\alpha$. It is easy to get
$h_{\alpha:c}=\langle a,b|N_{\alpha}|a,b\rangle=\alpha c+(1-\alpha)\langle a,b|L|a,b\rangle$,
which implies
$\langle a,b|L|a,b\rangle=(h_{\alpha:c}-\alpha c)/(1-\alpha)$.
Then, by the definition of $p_c(L)$, we have
$(h_{\alpha:c}-\alpha c)/(1-\alpha)\le p_c(L)$.
Thus, $h_{\alpha:c}\le \alpha c+(1-\alpha)p_c(L)$,
which, together with  (\ref{halpha}),  yields the conclusion (\ref{conclusion}). $\hfill \Box$
\\

Define
\[
V_{\alpha:c}=(\alpha c+(1-\alpha)p_c(L))I-N_{\alpha}.
\]
If the optimal state in $\mathcal{S}_{sep}$ to the test operator $N_{\alpha}$ belongs to $\mathcal{S}_{sep:\tilde c}$, then from Lemma 3.1 we can get
 $W_c(N_{\alpha})=V_{\alpha:c}$. Thus, $V_{\alpha:c}$ is an UEW.

Now, a natural question is what happens as $\alpha\rightarrow -\infty$. In fact, it is easy to get
\begin{align}
\label{limit1} \lim\limits_{\alpha\rightarrow -\infty}\left(\frac{1}{-\alpha}V_{\alpha:c}\right)&=\lim\limits_{\alpha\rightarrow -\infty} \left(\left(-c+\left(1-\frac{1}{\alpha}\right)p_c(L)\right)I+\left(C+\left(\frac{1}{\alpha}-1\right)L\right)\right)\\
\label{limit2} &=(p_c(L)- c)I-(L -C).
\end{align}
Thus, if $L-C$ is chosen to be a test operator, we can get the following ultrafine entanglement witness.\\
\\
 \textbf{$-\infty$-UEW: }\emph{For a given constraint value $c$, a given state $\rho$ is entangled if
\begin{align}
\label{uew2}\emph{Tr}(C\rho)\le c \wedge \emph{Tr}(W_c(L-C)\rho)<0\text{, or, } \emph{Tr}(C\rho)\ge c \wedge \emph{Tr}(W_{\tilde{c}}(L-C)\rho)<0.
\end{align}}

Hence, if the optimal state in $\mathcal{S}_{sep}$ to the test operator $L-C$ belongs to $\mathcal{S}_{sep:\tilde c}$, then from Lemma 3.1 and (\ref{limit1}) - (\ref{limit2}) we get the UEW
$$W_c(L-C)=(p_c(L)-c)I-(L-C).$$
\textbf{Lemma 3.2.} \emph{For a given constraint value $c$ and  real parameters $\alpha_1\le \alpha_2<1$,  suppose that the optimal pure state  in $\mathcal{S}_{sep}$ to the test operator $L$ belongs to $\mathcal{S}_{sep:\tilde c}$. If a given state $\rho$ satisfies $\emph{Tr}(\rho C)\le c, \emph{Tr}(V_{\alpha_{2}:c}\rho)<0$, then  $\emph{Tr}(V_{\alpha_{1}:c}\rho)<0$.}
\\
\\
\textbf{Proof.} Set
$\bar L=\alpha_2 C+(1-\alpha_2)L$ and ${\bar p}_c=\alpha_2 c+(1-\alpha_2)p_c(L)$.
Then $V_{\alpha_2:c}$ can be represented as $V_{\alpha_2:c}={\bar p}_cI-\bar L$.
Here, $\bar L$ can be regarded as a new test operator. Denote $\beta=\frac{\alpha_1-\alpha_2}{1-\alpha_2}$. $V_{\alpha_1:c}$ can be decomposed into
\begin{align}
V_{\alpha_1:c}&=(\beta c+(1-\beta){\bar p}_c)I-(\beta C+(1-\beta)\bar L)\\
&=\beta (cI-C)+(1-\beta)({\bar p}_cI-\bar L).
\end{align}
Thus, for any state $\rho$ with
$\text{Tr}(C\rho)\le c, \text{Tr}(V_{\alpha_2:c}\rho)<0$, i.e., $\text{Tr}((cI-C)\rho)\ge 0, \text{Tr}(({\bar p}_cI-\bar L)\rho)<0$, we get
$\text{Tr}(V_{\alpha_1:c}\rho)=\beta \text{Tr}((cI-C)\rho)+(1-\beta)\text{Tr}(({\bar p}_cI-\bar L)\rho)<0$,
where we have used $\beta\le 0$. $\hfill \Box$
\\

We now give comparisons among (\ref{uew}), $\alpha$-UEWs and $-\infty$-UEW.
\\
\\
\textbf{Proposition 3.1.} \emph{For a given constraint value $c$, if the optimal states in $\mathcal{S}_{sep}$ to the test operators $L$ and $L-C$ both belong to $\mathcal{S}_{sep:\tilde c}$, then for any $\alpha<1$ we have the UEW $W_c(N_{\alpha})=V_{\alpha:c}$, which is more efficient when $\alpha$ gets smaller. Furthermore, the $-\infty$-UEW
$$W_c(L-C)=(p_c(L)-c)I-(L-C)$$
 is the optimal UEW among $V_{\alpha:c}$ for any $\alpha<1$.}
\\
\\
\textbf{Proof.} We first prove that the optimal state in $\mathcal{S}_{sep}$ to the test operator $N_{\alpha}$ for any $\alpha<1$ belongs to $\mathcal{S}_{sep:\tilde c}$. Otherwise, there exists a separable state $\delta_{sep}\in \mathcal{S}_{sep:c}$ such that $\text{Tr}(V_{\alpha:c}\delta_{sep})<0$. From Lemma 3.2 and  (\ref{limit1})-(\ref{limit2}), we can obtain $\text{Tr}(W_c(L-C)\delta_{sep})<0$, which contradicts to the fact that the optimal state in $\mathcal{S}_{sep}$ to the test operator $L-C$ belongs to $\mathcal{S}_{sep:\tilde c}$. Hence, from Lemma 3.1, we can get $W_c(N_{\alpha})=V_{\alpha:c}$. Due to Lemma 3.2, $V_{\alpha:c}$ is more efficient when $\alpha$ gets smaller. Thus, from (\ref{limit1})-(\ref{limit2}), the $-\infty$-UEW $W_c(L-C)=(p_c(L)-c)I-(L-C)$ is the optimal UEW among $V_{\alpha:c}$ for any $\alpha<1$. $\hfill \Box$
\\

We now consider the case that the optimal states in $\mathcal{S}_{sep}$ to the test operators $L$ and $L-C$ belong to $\mathcal{S}_{sep:\tilde c}$ and $\mathcal{S}_{sep:c}$, respectively.
\\
\\
\textbf{Proposition 3.2.} \emph{For a given constraint value $c$ and
\[
\alpha_0=\inf\left\{\alpha|1>\alpha\in \mathbb{R},\emph{Tr}(V_{\alpha:c}\rho)\ge 0,\forall \rho\in \mathcal{S}_{sep:c}\right\},
\]
if the optimal states in $\mathcal{S}_{sep}$ to the test operators $L$ and $L-C$ belong to $\mathcal{S}_{sep:\tilde c}$ and $\mathcal{S}_{sep:c}$, respectively, then $W_c(N_{\alpha_0})=V_{\alpha_0:c}$ is a FEW.
}
\\
\\
\textbf{Proof.} From Lemma 3.2, it only needs to show $\text{Tr}(V_{\alpha_0:c}\rho_{sep})\ge 0$ for any separable state $\rho_{sep}$. We consider the following two cases.

\emph{Case} (i): $ \rho_{sep}\in \mathcal{S}_{sep:c}$. It is trivial from the definition of $\alpha_0$.

\emph{Case} (ii): $ \rho_{sep}\in \mathcal{S}_{sep:\tilde{c}}$. If $\text{Tr}(V_{\alpha_0:c}\rho_{sep})<0,$ then the optimal state in $\mathcal{S}_{sep}$ to $N_{\alpha_0}$ belongs to $S_{sep:\tilde c}$. Thus, there must exist $\alpha_1<\alpha_0$ such that $\text{Tr}(W_c(N_{\alpha_1})\rho_{sep})=0$. Thus, from Lemma 3.1 the optimal state in $\mathcal{S}_{sep}$ to the test operator $N_{\alpha_1}$ also belongs to $\mathcal{S}_{sep:\tilde c}$. Furthermore, we have $\text{Tr}(V_{\alpha_1:c}\rho)\ge 0, \forall \rho\in \mathcal{S}_{sep:c}.$ This contradicts the definition of $\alpha_0$. $\hfill \Box$
 \begin{figure}[htbp]
\small
\centering
 \subfigure[\emph{The optimal state in $\mathcal{S}_{sep}$ to the test operator $L-C$ belongs to $\mathcal{S}_{\tilde c}$.
 }]{
\begin{minipage}[b]{0.4\textwidth}
\centering
\includegraphics[width=2.3in]{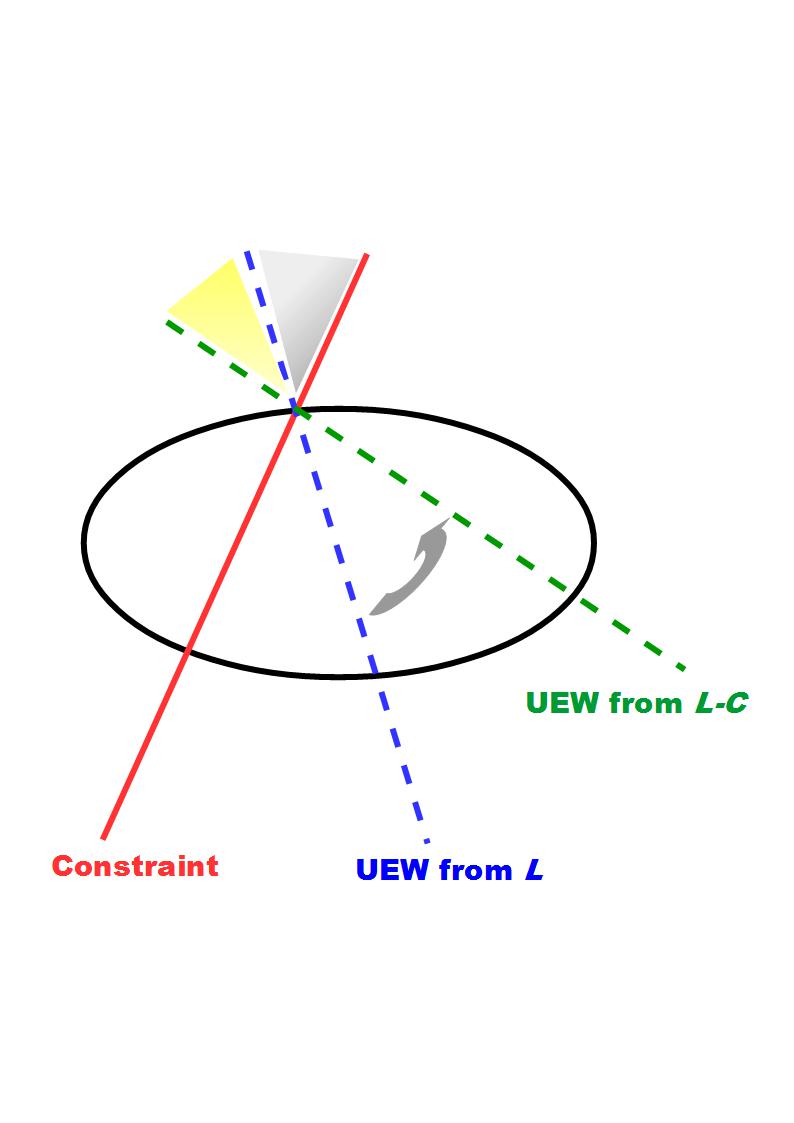}
\end{minipage}}
\subfigure[\emph{The optimal state in $\mathcal{S}_{sep}$ to the test operator $L-C$ belongs to $\mathcal{S}_{c}$. }]{
\begin{minipage}[b]{0.4\textwidth}
\centering
\includegraphics[width=2.3in]{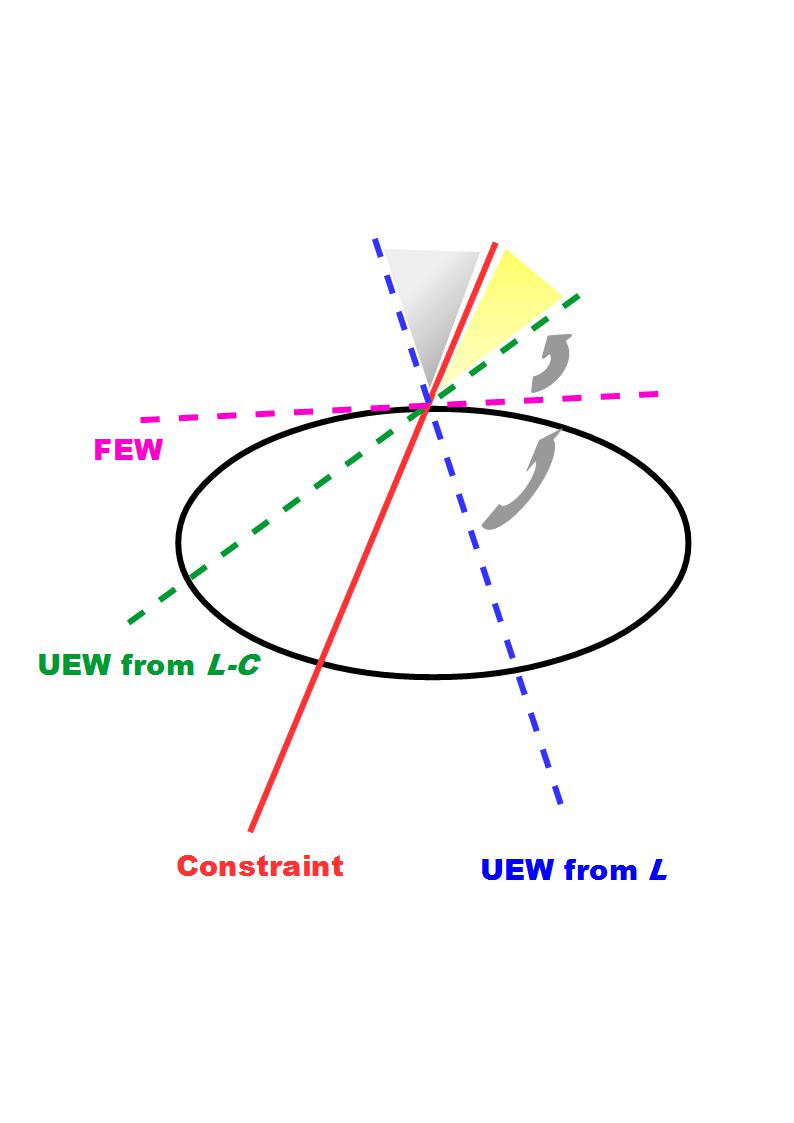}
\end{minipage}}
\caption{(Colour figure) \emph{Perceptional illustrations of rotation and optimization.} In Fig. (a), the optimal UEW $W_c(L-C)$ (the green broken line) can detect the entangled states located in the yellow and gray regions, but $W_c(L)$ (the blue broken line) cannot detect the states in the yellow region. In Fig. (b), the finest EW (the pink broken line) is the optimal UEW, that can detect all states above the pink broken line. The UEW $W_{\tilde{c}}(L-C)$ (the green broken line) can detect entangled states located in the yellow region, while $W_c(L)$ (the blue broken line) can only detect the states in the gray region.}
\end{figure}
\\

 Based on Lemmas 3.1-3.2 and Propositions 3.1-3.2, we have the following conclusions.
\\

\emph{For a given constraint value $c$ and a real parameter $\alpha<1$,  suppose that the optimal state  in $\mathcal{S}_{sep}$ to the test operator $L$ belongs to $\mathcal{S}_{sep:\tilde c}$, then the following results hold:
\begin{description}
  \item[Case (i):] The optimal state  in $\mathcal{S}_{sep}$ to the test operator $L-C$ belongs to $\mathcal{S}_{sep:\tilde c}$.
      In this case, the UEW $W_c(N_{\alpha})=V_{\alpha:c}$  becomes more efficient when $\alpha$ gets smaller from $1$ to $-\infty$. Thus, the $-\infty$-UEW $W_c(L-C)$ is the optimal one among all $V_{\alpha:c}$. In particular, $W_c(L)$ is less efficient than $V_{\alpha:c}$ for all $\alpha<0$. See Fig. \emph{1(}a\emph{)} for a perceptional illustration.
      \item[Case (ii):] The optimal state in $\mathcal{S}_{sep}$ to the test operator $L-C$ belongs to $\mathcal{S}_{sep:c}$. In this case, the UEW $W_c(N_{\alpha})=V_{\alpha:c}$ becomes more efficient when $\alpha$ gets smaller from $1$ to $\alpha_0$, and the UEW $W_{\tilde c}(N_{\alpha})$ becomes more inefficient when $\alpha$ gets smaller from $\alpha_0$ to $-\infty$. Thus, the finest entanglement witness $V_{\alpha_0:c}$ is the optimal one among these ultrafine entanglement witnesses. In particular, $W_c(L)$ is less efficient than $V_{\alpha:c}$ for any $\alpha\in [\alpha_0,0).$ The UEW $W_{\tilde c}(L-C)$ can detect some entangled states that $V_{\alpha:c}$ cannot for any $\alpha\in (\alpha_0,0]$. See Fig. \emph{1(}b\emph{)} for a perceptional illustration.
\end{description}
}
In Section 1, four merits are summarized for UEWs $W_c(L)$ and $W_{\tilde c}(L)$. Clearly, the $\alpha$-UEWs have the same merits. Moreover, the $\alpha$-UEWs can also be extended to the multipartite scenario by similar arguments in \cite{Shahandeh2017}.
We now give an example to further illustrate Case (i).
\\
\\
\textbf{Example 3.1.} We use the following two qubit pure state derived from \cite{Shahandeh2017}:
\[
|\phi\rangle=\alpha|00\rangle+\beta|01\rangle+\beta|10\rangle+\delta|11\rangle,
\]
where $\alpha={7}/{10}$, $\beta={1}/{2}$ and $\delta=\sqrt{1-\alpha^2-2\beta^2}$.
Consider the mixture of the state $|\phi\rangle$ with the white noise
\[
\rho_p=\frac{p}{4}I+(1-p)|\phi\rangle\langle \phi|,
\]
where $0\le p\le 1$. By using positive operator-valued measure (POVM) elements
\[
P_1=x|1\rangle \langle 1|,~~P_2=|\xi^+\rangle\langle \xi^+|,~~P_3=|\xi^-\rangle\langle \xi^-|
\]
with
\[
|\xi^{\pm}\rangle=\frac{1}{\sqrt{2}}|1\rangle\pm\sqrt{\frac{1-x}{2}}|0\rangle,
\]
from \cite{Shahandeh2017} we have the constraint and test operators
$C=P_1\otimes P_1$ and $L=P_2 \otimes P_2$, respectively. If the constraint value $c$ and the parameter $x$ are respectively chosen to be $\frac{1}{100}$ and $\frac{2}{3}$, then,  by MATHEMATICA computations, the optimal states in $\mathcal{S}_{sep}$ to the test operators $L$ and $L-C$ both belong to $\mathcal{S}_{sep:c}$. Table 1 gives entanglement conditions of $\rho_p$ from $\alpha$-UEWs.

\begin{table}[htbp]
\centering \begin{tabular} {c|c|c|c|c}\hline
{$\alpha=0$}&$\alpha=-1$&$\alpha=-10$ &$\alpha=-100$&$\alpha=-\infty$
 \\ \hline
 {--}&$0\le p\le 0.004$ & $0\le p\le 0.009$& $0\le p\le 0.010$& $0\le p\le 0.010$
\\
 \hline
\end{tabular} \caption{\emph{Entanglement conditions of  $\rho_p$ from the $\alpha$-UEW $W_{c}\left({N_{\alpha}}\right)$ with different values of $\alpha$. The symbol ``${-}$" denotes that the UEW $W_{c}\left({N_{0}}\right)=W_c(L)$ cannot detect any entanglement in $\rho_p$. For the case $\alpha=-\infty$, the UEW $W_{c}\left({N_{\alpha}}\right)$ reduces to the $-\infty$-UEW $W_{c}(L-C)$.
}}
\label{tab}
\end{table}

It can be seen from Table \ref{tab} that the $\alpha$-UEW $W_{c}(N_{\alpha})$ is more efficient when $\alpha$ gets smaller from $0$ to $-\infty$. Thus, the UEW $W_c(L-C)$ is the optimal one among all UEWs $W_{c}(N_{\alpha})$ with different values of $\alpha$. This coincides with our theoretical results.

\section{Conclusions}

By rotating the hyperplane determined by the test operators, a series of UEWs have been derived. The optimal UEWs among these UEWs have been also investigated.
These UEWs not only have the same advantage as the UEW from $L$, but also can detect some entangled states that the UEW from $L$ cannot detect. Nevertheless, a simply linear combination of the test and the constraint operators is only used in this paper. How to optimize the original UEW by some more innovative methods is still an interesting problem.

\section*{\bf Acknowledgments}
The authors greatly indebted to the referee and the editor for their comments and suggestions. This work is supported by the Natural Science Foundation of Shandong Province (ZR2016AM23, ZR2016AQ06), the Fundamental Research Funds for the Central Universities (18CX02035A, 16CX02049A) and the NSF of China (11675113, 11775306).

\end{document}